# Non-linear X-ray Coherent Diffractive Imaging

Arnab Sarkar and Allan S Johnson[*]

*IMDEA Nanoscience, Calle Faraday 9, 28023 Madrid, Spain*

**ABSTRACT.** The advent of nonlinear X-ray processes like sum-frequency generation and four-wave mixing raises the possibility of non-linear X-ray imaging, combining the high-resolution and elemental specificity of X-ray imaging with the state selectivity and sensitivity of non-linear optical imaging. While scanning imaging methods may be feasible, for linear X-ray processes coherent diffractive imaging has emerged as a key approach, enabling lensless reconstruction of nanoscale structure and dynamics with high spatial and temporal resolution. In this work, we propose a coherent diffractive approach to imaging using X-ray nonlinear processes, introducing an analysis method to isolate the nonlinear component from the overall diffraction pattern by leveraging the property of mutual incoherence between different wavelengths. For examples such as sum-frequency generation in ferroelectrics, this method reveals both domain structure and orientation through the retrieved amplitude and phase of the nonlinear signal. We discuss the feasibility of the proposed method in the presence of experimental noise, most relevantly shot noise. This analysis method is applicable in both static and dynamic imaging, offering a pathway beyond traditional spectroscopy toward XUV/X-ray coherent imaging of spatio-temporal dynamics in quantum materials and biological systems.

## I. INTRODUCTION

X-rays are widely used as a noninvasive probe to study materials and structures across many scientific disciplines and interdisciplinary fields, often with electronic and structural information. Because of their short wavelengths and high photon energies, X-rays provide high resolution, large penetration depth, and elemental sensitivity for diffraction, coherent imaging, and spectroscopic measurements [1,2]. Meanwhile, in the optical domain, nonlinear light–matter interactions have been widely employed to investigate crystal symmetry, surface and interface properties, phase-contrast imaging of non-centrosymmetric structures, and label-free visualization of biological systems [3]. For example, sum-frequency generation (SFG) microscopy has proven to be a powerful tool for imaging ferroelectric domain structures and domain boundaries [4–7].

Recent advances in synchrotron radiation and X-ray free-electron lasers (XFELs) have enabled the experimental observation of such nonlinear phenomena with X-rays [8–12], including X-ray second harmonic generation (SHG) [13–15], sum and difference frequency generation [16,17], and three- and four-wave mixing processes involving X-ray and optical fields [18–21], thereby raising the possibility of nonlinear X-ray imaging - combining the high resolution of X-rays with the selectivity of nonlinear optics. Transient resonances have been found to enhance X-ray scattering cross sections [22], but thusfar there has been little discussion regarding the possibility of more conventional wavemixing-type non-linear X-ray imaging to date.

In the simplest implementation of non-linear X-ray imaging a relatively long focal length optic would focus X-ray to a high-intensity point on the sample, and the emitted radiation spectrally and/or spatially filtered to isolate the non-linear signal. The sample is then scanned to build up a real space image. Long focal lengths are necessary to protect the x-ray optics but also to enable access for mixing with optical fields, as X-ray+optical wavemixing is orders of magnitude more efficient than pure X-ray mixing and thus the most promising candidate for nonlinear X-ray wavemixing. However, this comes at a significant drawback in terms of spatial resolution and in high stability requirements which are difficult to meet with high-intensity X-ray sources, and indeed for conventional linear X-ray imaging it is common to instead use lensless imaging methods [23] like Fourier Transform Holography (FTH) [24], Coherent Diffraction Imaging (CDI) [25,26] or Ptychography [27,28]. These methods rely on capturing the X-ray scatter from a sample, and can achieve diffraction limited imaging with detectors placed at long distances from the samples. Coherent methods have been heavily leveraged with intense coherent X-ray sources [23] and can be combined with high-fluence optical excitation [1].

------

[*]Contact author: allan.johnson@imdea.org

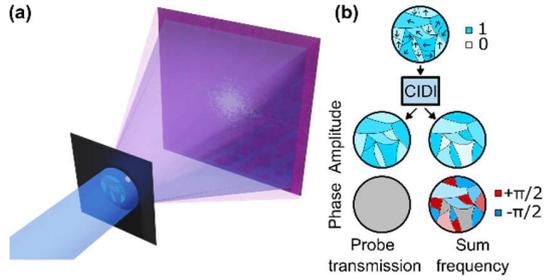

*Fig1. Schematic diagram:* (a) A schematic diagram of coherent diffraction imaging experiment is shown where probe beam is represented in blue. Diffracted signal with additional frequency component (represented in purple) is detected together on a 2D detector. (b) Schematic representation of a ferroelectric material showing inhomogeneous absorption of X-ray in nanoscale domains. Multimodal or CIDI analysis separates the absorption and the SFG signal where information of the orientation of the domains is imprinted in the phase component.

Here we demonstrate a scheme for implementing non-linear X-ray coherent diffractive imaging. Since conventional spectral filtering is impractical in the X-ray regime and spatial filtering schemes (i.e. gratings) are incompatible with CDI geometry, we instead explore post-processing strategies that analytically or numerically disentangle wavelength components from the recorded diffraction pattern. In particular we consider the application of multimodal-CDI (MMCDI) to recover both fundamental and nonlinear components, analogous to recent advances in using multi-colour X-ray probes for CDI [29–31], as well at the application of the analytic Coherence Isolated Diffractive Imaging (CIDI) technique [32]. We show that, for realistic conversion efficiencies of the nonlinear signal, CIDI successfully extracts the nonlinear contribution from coherent diffraction patterns, whereas MMCDI fails to recover it. We assess the feasibility of nonlinear X-ray coherent imaging in light of the current experimental state of the art and, using ferroelectric domain imaging as a model case, demonstrate that the CIDI approach is viable with existing sources and achievable within reasonable acquisition times.

## II. COHERENT IMAGING APPROACH

A typical XUV/Soft X-ray coherent imaging/FTH geometry is shown in Fig 1a, where a coherent beam (blue) is diffracted through a thin sample in a finite mask. A phase reference is acquired from the transmitted beam through small reference holes on the opaque support structure. The diffracted beam through the semi-transparent sample interferes with the reference beam and the diffracted intensity pattern is recorded on a two-dimensional detector. If the field strength is high enough or the beam is mixed with a high-field auxiliary optical beam, new nonlinear frequency component can be generated in the sample, which also scatter to the detector plane. Thus, the diffracted signal contains additional frequency components generated due to nonlinear interaction between the X-rays and sample. To apply conventional imaging approaches, the newly generated wavelength must be separated from the probe beam. However, since preserving the wavevector geometry of the diffracted beam is essential, as the resolution achievable related to the highest transverse wavevector captured, the grating based-spectrometers currently used in non-linear imaging experiments are not suitable for this task. Spectral filtering is perhaps even more challenging; by using thin films with different atomic constituents, different spectral regions can be enhanced and suppressed, but these regions are usually 10s to 1000s of electronvolts separate in photon energy, and are not suitable for separation wavelengths only a few eV apart. We briefly note that while such an approach may be suitable for second-harmonic generation (SHG), recent experiments have shown that even with resonant attosecond XUV pulses the intensity required to obtain observable SHG signals lead to sample damage in $TiO_2$ [15]. As such SFG between optical/IR and X-ray wavelengths appears to be the most feasible wavelength range for nonlinear X-ray imaging.

As physically separating the original and nonlinear frequency components is infeasible, we instead consider numerical data analysis approaches in keeping with the spirit of coherent imaging methods. The key factor is that the nonlinear component is mutually incoherent with the diffracted probe beam and doesn't interfere with the reference beam. As a result, an incoherent sum of the components is recorded on the detector. One approach to reconstructing multiple mutually incoherent components is MMCDI, in which the algorithm employs multiple support masks, each scaled to the corresponding wavelength, to simultaneously retrieve the probe field and the nonlinear components. While MMCDI can successfully retrieve signals at different wavelengths when their amplitudes are comparable (see supplementary material); however, under practical experimental parameters we will show it fails to recover the nonlinear component, even in the absence of shot noise, highlighting the need for a more efficient method to extract the nonlinear (SFG) pattern.

*Contact author: allan.johnson@imdea.org

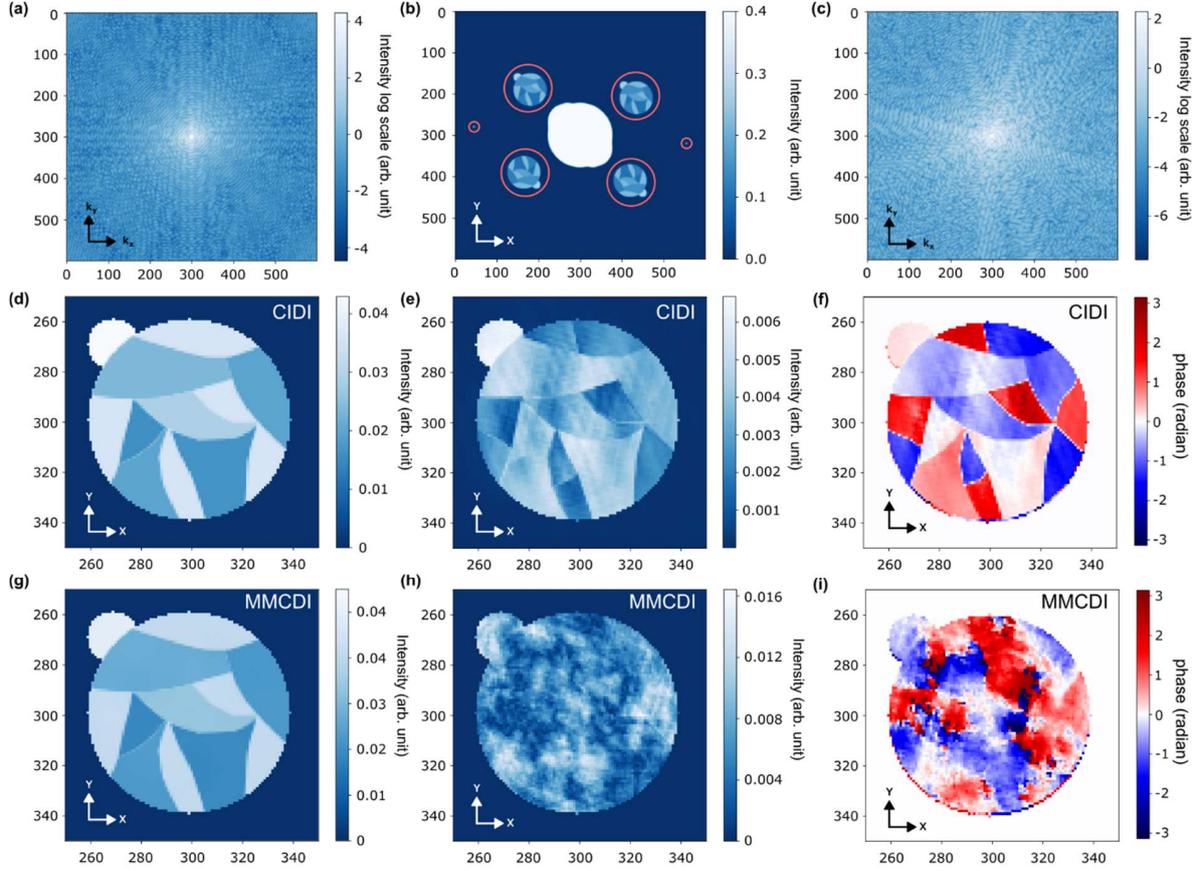

FIG 2. CIDI and multimodal analysis of SFG: (a) Total diffraction pattern: the incoherent sum of diffraction pattern of linear and nonlinear (SFG) components, (b) hologram of the diffraction pattern with all the cross-correlation terms highlighted in red circles and (c) isolated diffraction pattern off the SHG component using CIDI is generated using the sample image shown in figure 1b. CIDI reconstruction of (d) amplitude of the linear component, (e) amplitude, and (f) phase of the SFG component for realistic conversion ratio without noise is shown. For the same parameters, MMCDI can retrieve the (g) amplitude of the linear component, but unable to retrieve the (h) amplitude, and (i) phase of the SFG component.

We have recently introduced an alternative methodology, CIDI, which can separate out mutually incoherent components in a phase-referenced diffraction pattern [32]. Originally introduced for examining fluctuations, as it does not require that the second pattern is self–coherent as in MM CDI, here we show it applies equally well self-coherent non-linear signals. We briefly recap the salient points of the approach here. Let us consider that $E(r)$ is the total electric field after the beam is transmitted through sample and reference holes. The electric field can be expressed as:

$$E(r) = x(r) + s(r) + \sum R(r).$$

Here, $x(r)$, and $s(r)$, are probe and SFG component from the sample and $R(r)$ is the contribution from the reference holes.

The diffraction pattern on the detector is:

$$|f(k,t)| = |FFT(E(r))|^2$$
$$= |x(k)|^2 + |s(k)|^2 + \sum |R(k)|^2$$
$$+ \sum R(k)x^*(k) + \sum R^*(k)x(k)$$
$$+ \sum\sum R_i^* R_j$$

where the cross-correlation terms between s(k) and the other components have been set to zero due to their mutual incoherence, The autocorrelation term of the nonlinear component $|s(k)|$ can then be expressed as:
$|s(k)|^2 = |f(k,t)|^2 - |x(k)|^2 - |R(k)|^2 -$
$\sum R(k)x^*(k) - \sum R^*(k)x(k).$ (1)

The autocorrelation terms in the right-hand side can be expressed in terms of the cross-correlation terms (see supplementary info for details). Using this expression, incoherent and coherent part can be isolated from each other and CDI applied on the separated diffraction patterns individually [32].

---

*Contact author: allan.johnson@imdea.org

### III. RESULTS

We consider SFG from a ferroelectric material with domains of different orientations, as shown in Fig. 1b. SFG, the lowest order X-ray nonlinear interaction, is considered for two major reasons: (i) Maximum conversion efficiency among all other nonlinear X-ray processes and (ii) the local non-centrosymmetric nature is imprinted in the phase of the generated second harmonic, but is lost in linear imaging. In $\chi_2$ processes, ferroelectric domains of opposite polarity generate the same sum-frequency intensity but with opposite phase. Fig 1b shows local absorption and orientation of the ferroelectric domains considered, along with the corresponding amplitude and phase of probe transmission and SFG. The transmitted intensity is identical for domains of the same orientation but opposite polarity, but the phase of the generated SFG is inverted for opposite polarities.

For numerical analysis, we assume the SFG amplitude follows the same distribution as the transmission image. The sample is masked by opaque circular support and the symmetry of the support mask is broken by adding a small structure at one side of the support mask to achieve better convergence during CDI iterations. At least two reference holes are required to apply CIDI on the recorded diffraction pattern [32]. For clean separation of the terms in eq. (1), the distance between the sample and transmissive reference holes must exceed $2D$, where $D$ is the diameter of the circular support mask. In the presence of nonlinear components, the diffraction pattern scales with the wavelength, and the support mask diameter scales accordingly. Thus, for CIDI, the distance must be greater than $2D$, with $D$ defined by the diameter of the largest support mask corresponding to the lowest detected wavelength. Figure 2b shows the total diffraction pattern incoherently summing the contributions and considering a 2% SFG conversion efficiency, while Figure 2c shows the Fourier transform of the diffraction pattern. Four side-band terms (holographic images) are clearly visible around the autocorrelation term, along with the cross-correlation terms of the two references indicated by red circles. Applying the CIDI algorithm leads to the isolated intensity diffraction pattern of the SFG component shown in figure 2c.

Figures 2d-f show the reconstructed linear transmission, nonlinear amplitude and nonlinear phase respectively, for the CIDI approach, while figures 2g-i show the same for the MMCDI approach. Interestingly, though the SHG amplitudes are the same, but a domain boundary appears in the reconstruction in Figure 2e because the sharp phase jump introduces a discontinuity in the complex field, which appears as a high-frequency component that manifests as an amplitude edge. The CIDI phase reconstruction closely matches the input image (figure 2b), whereas MMCDI fails to recover both amplitude and phase of the nonlinear component. This demonstrates that CIDI can isolate probe transmission and the SFG signal from a single dataset and reconstruct both simultaneously, while MMCDI cannot recover the nonlinear part for such small conversion efficiencies (Supplementary information).

### IV. EXPERIMENTAL FEASIBILITY

Having demonstrated the ability of CIDI to recover nonlinear coherent signals, we next examine the experimental feasibility of the approach under realistic conversion efficiencies and signal-to-noise conditions. The low conversion efficiency of the XUV/X-ray nonlinear processes presents the biggest challenge to image them in presence of different kind of noise in experimental setups, and estimating these cross-sections is currently challenging. In a recent experiment, the conversion efficiency of extreme-ultraviolet SHG was reported to be 0.02, though in this case the material ($TiO_2$) was observed to suffer damage and had to be rastered to new points continuously [15]. Optical-X-ray SFG may be more efficient, but around 2% would seem a reasonable estimate. Conversion efficiencies of other X-ray nonlinear processes are generally orders of magnitude lower than SFG or SHG. For instance, recent spectroscopy experiments of four wave mixing of XUV/X-ray in materials have shown that the intensity ratio between input XUV and generated FWM peak intensity is limited 0.01% around, with the value of $\chi_3$ on the order of $10^{-1}$ $cm^2/W$ [18–21].

While overall intensity fluctuations of the X-rays do not adversely the imaging due to the full field-of-view nature of the coherent imaging [2], this is not true for shot and read noise, which can be expected to significantly impact the ability to perform reconstructions when considering such low conversion efficiencies. Shot-noise is particularly challenging as it scales with signal level, and as the brightest parts of the diffraction pattern can be several orders of magnitude higher than the signal in the weakest, it can dramatically complicate the reconstruction processes.

______________

*Contact author: allan.johnson@imdea.org

As the signal-to-noise (SNR) of shot noise scales as $\sqrt{n}$, where n is the number of photons, in principle this contribution can be reduced simply by longer acquisitions. This can practically be achieved either by increasing exposure times or by averaging over multiple acquisitions. However, the exposure time has an upper limit when the detector is saturated. Commercially available two-dimensional CCD cameras for XUV or soft X-ray detection typically have well-depths of $2^{18} = 262144$ electrons. As the activation energy per electron in such detectors is around 3.7 eV, this means for 50 eV photons only ~20000 photons can be detected per pixel before saturation (see supplementary material). At this level the shot-noise level is comparable to the nonlinear signal, and averaging across many nearly saturated exposures becomes essential.

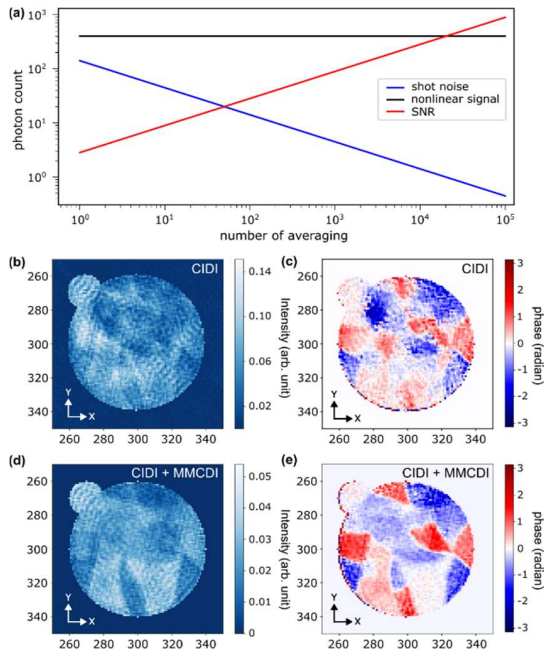

**FIG 3. Effect of noise:** *(a) Considering conversion efficiency: 0.02, and maximum possible photon count on the detector before saturation: 20000, the shot noise due to the linear component of a single shot image becomes comparable to the SFG component. Averaging over multiple snapshots reduces the shot noise level and improves the signal to noise ratio (SNR) significantly for good CIDI reconstruction. In presence of shot noise, CIDI is performed on dataset averaged over $5*10^3$ snapshots to get reconstruction of (b) amplitude, and (c) phase of the SFG component. MMCDI on CIDI-isolated diffraction pattern improves both (d) amplitude and (e) phase by separating the signal and noise in two different support masks.*

In Fig. 3a, we investigate the dependence of shot noise and signal-to-noise ratio (SNR) on the number of averaged images by plotting the mean signal (black line), shot noise (blue line), and SNR (red line) of the maximum camera pixel, with the exposure set to match the well-depth at the brightest pixel. This plot provides an estimate of the minimum number of images required for reliable reconstruction, though it reflects only the maximum pixel. According to this estimate for 2% conversion efficiency, signal should be retrievable from one single image; in practice, the situation is much worse as significant signal is required at the weakest pixels. We find at least $5*10^3$ images must be averaged because the reconstruction methods correlate signals at both the highest and lowest signal level regions of the detector. This requirement can be reduced by employing a Gaussian absorption beam stop, which minimizes the contrast between the highest and lowest signal levels [33]. Figure 3b and 3c show the reconstruction of SHG amplitude, and SHG phase from averaged diffraction pattern in the presence of shot and read noise (2.5 electrons/pixel) for $5*10^3$ averaged acquisitions. Assuming around 1s exposures [34], this leads to around 1.5 hours of acquisition to take a nonlinear image.

Since CIDI isolates all components incoherent to the reference signal, the noise naturally appears in the incoherent part together with the nonlinear signal. A hybrid scheme applying MMCDI to the CIDI-isolated diffraction pattern enables separation of noise and nonlinear signals into distinct support masks, thereby significantly improving both amplitude and phase reconstructions. The MMCDI outputs based on the CIDI- isolated diffraction pattern are presented in Fig. 3d and 3e, respectively. Further improvements may be possible by leveraging schemes for the separation of shot-noise from conventional X-ray imaging [35,36] or by acquiring reference images with optical fields to aid in the isolation of the linear component, though optical modification of the linear X-ray properties will in general be difficult to discard at such high field strengths.

## V. CONCLUSIONS

In summary, we propose an experimental strategy to image X-ray nonlinear processes through coherent diffraction imaging and numerically demonstrate that CIDI can isolate these signals. This capability extends beyond spectroscopy by enabling access to nanoscale nonlinear interactions within materials. The method is versatile across static and ultrafast regimes, and our analysis addresses the feasibility of experiments in which nonlinear signals are orders of magnitude weaker than the probe intensity. At the current state of the art, overcoming shot noise and read noise requires averaging over many exposures. Employing low-noise electronics and cooling the detector effectively reduces its impact, and the acquisition times predicted are well within experimental viability. This work takes us a step beyond conventional spectroscopy by enabling direct imaging of X-ray nonlinear processes, andopens the door to XUV/X-ray nonlinear coherent imaging on ultrafast timescales, providing access to the spatio-temporal dynamics of nonlinear interactions at the nanoscale.

*Contact author: allan.johnson@imdea.org

**ACKNOWLEDGMENTS**

This work was funded by the Spanish AIE (projects PID2022-137817NA-I00 and EUR2022-134052) and the Comunidad de Madrid (project TEC-2024/TEC-380 "Mag4TIC"), and with the support of a 2024 Leonardo Grant for Scientific Research and Cultural Creation, BBVA Foundation. Funded by the European Union (ERC, KnotSeen, 101163311). Views and opinions expressed are however those of the author(s) only and do not necessarily reflect those of the European Union or the European Research Council. Neither the European Union nor the granting authority can be held responsible for them. ASJ acknowledges the support of the Ramón y Cajal Program (Grant RYC2021-032392-I). IMDEA Nanociencia acknowledges support from the "Severo Ochoa" Programme for Centers of Excellence in R&D (MICIN, CEX2020-001039-S).

*Contact author: allan.johnson@imdea.org

---

*Contact author: allan.johnson@imdea.org

# Supplementary material: Non-linear X-ray Coherent Diffractive Imaging

Arnab Sarkar and Allan S Johnson*

*IMDEA Nanoscience, Calle Faraday 9, 28023 Madrid, Spain*

## 1. Detail calculation of Coherence Isolated Diffractive Imaging

Let us consider $E(r)$ as the total electric field after the beam is transmitted through sample and reference holes. The electric field can be expressed as:

$$E(r) = x(r) + s(r) + \sum R(r).$$

Here, $x(r)$, and $s(r)$, are coherent and nonlinear contribution from the sample and $R(r)$ is the contribution from the reference holes. The diffraction pattern on the detector will be:

$$|f(k,t)| = \left|FFT\big(E(r)\big)\right|^2 = |x(k)|^2 + |s(k)|^2 + \sum|R(k)|^2 + \sum R(k)x^*(k) + \sum R^*(k)x(k) + \sum\sum R_i^* R_j.$$

To isolate $|x(k)|^2$ and $|s(k)|^2$ from the whole diffraction pattern, all the cross-correlation terms need to be calculated from the hologram. The final expressions of the components are:

$$|x(k)|^2 = \left|\sqrt{\frac{R_i(k)x^*(k) \times R_i^*(k)x(k) \times R_j(k)x^*(k) \times R_j^*(k)x(k)}{R_i^*(k)R_j(k) \times R_i(k)R_j^*(k)}}\right|;$$

for any $i, j$ when $i \neq j$, and

$$|s(k)|^2 = \left|FFT\big(E(r)\big)\right|^2 - |x(k)|^2 - |R(k)|^2 - \sum R(k)x^*(k) - \sum R^*(k)x(k).$$

In the last expression, autocorrelations of reference holes are calculated following the equation:

$$|R_i(k)|^2 = \frac{R_i(k)x^*(k) \times R_i^*(k)x(k)}{|x(k)|^2}.$$

Here, both $|x(k)|^2$ and $|s(k)|^2$ is expressed only in terms of the cross-correlations.

---

*Contact author: allan.johnson@imdea.org

## 2. CIDI vs MMCDI reconstruction of SFG under high conversion efficiency

When the SFG signal strength is approximately equal to the probe transmission (1:1), both CIDI and MMCDI are capable of accurately reconstructing the SFG amplitude and phase.

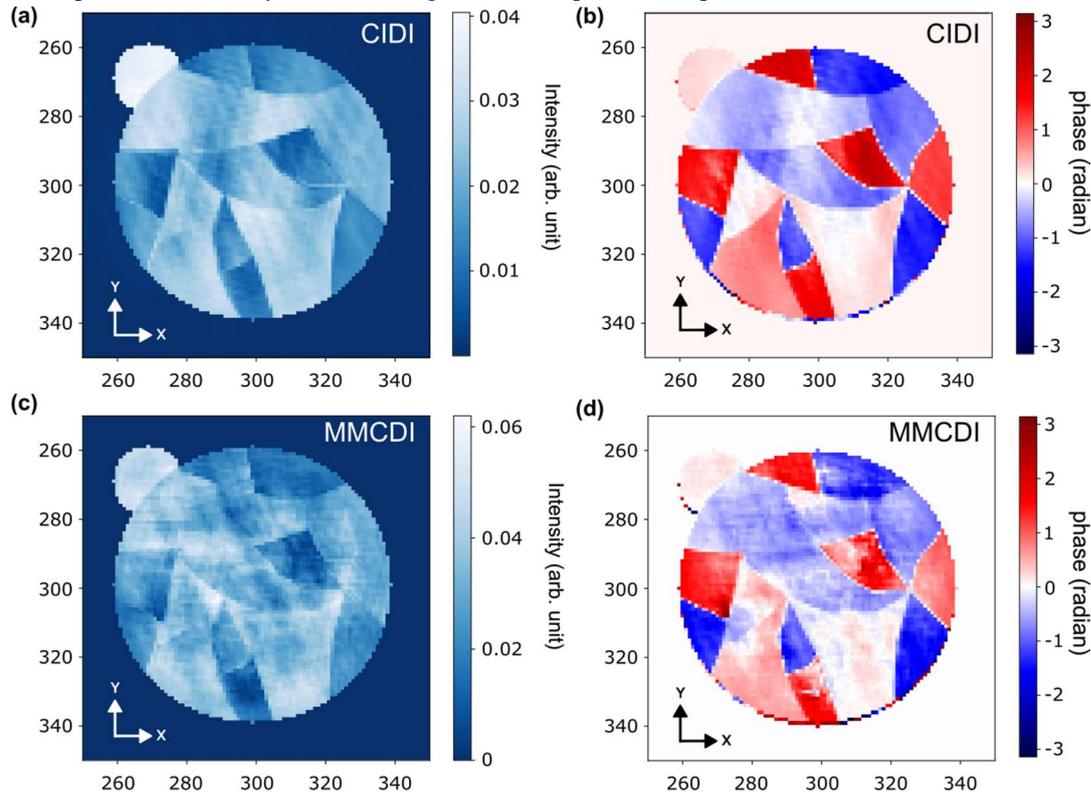

*Figure S1: CIDI reconstruction of (a) amplitude, and (b) phase of the SFG component for 50% conversion ratio and without noise is shown. For the same parameters, MMCDI also retrieves both (c) amplitude, and (d) phase of the SFG component.*

## 3. Parameters to understand the feasibility of the proposed experiment

To verify the experimental feasibility of nonlinear imaging, we introduce noise into the signal. The following parameters are considered for the simulations:

Wavelength of X-ray: 50 eV
Well depth/ADC precision: $2^{18}=262144$ electrons
Activation energy of one electron: 3.7 eV
Maximum possible photon number per pixel before saturation: ~20000 photons
Conversion efficiency: 0.02 (2%)
Read noise: 2.5 e- standard deviation. Minimum 0.

---

*Contact author: allan.johnson@imdea.org